\documentstyle[prl,floats,aps,twocolumn,epsf,graphicx]{revtex}
\begin{document}
\twocolumn[\hsize\textwidth\columnwidth\hsize\csname
@twocolumnfalse\endcsname

\title{Unruh radiation: from black holes to elementary particles}
\author{Shahar Hod}
\address{The Racah Institute of Physics, The Hebrew University, Jerusalem 91904, Israel}
\date{\today}
\maketitle

\begin{abstract}

\ \ \ We advocate the idea that Unruh's quantum radiation, whose theoretical discovery was originally 
motivated by the physics of black holes, may have important implications on the structure and dynamics 
of elementary particles. To that end, we analyze the Unruh radiation effect experienced by an 
accelerated particle in atomic and nuclear systems. 
For atomic systems, the effect is negligible as compared to the characteristic energy of the system. 
On the other hand, it is found that a quark inside a nucleon may experience Unruh radiation whose energy 
is of the same order of magnitude as the quark's own mass. 
\end{abstract}
\bigskip

]

One of the most remarkable theoretical predictions of the former century is Hawking's celebrated result 
that black holes should evaporate by the emission of {\it thermal} radiation \cite{Haw}. 
Subsequent to Hawking's discovery, Unruh \cite{Unruh1} realized that many features of the black-hole 
radiation could be understood in the simpler context of Minkowski spacetime. 
In particular, he found that the Minkowski vacuum, i.e., the quantum state associated with the 
nonexistence of particles according to an inertial observer, corresponds to a {\it thermal} bath 
of elementary particles at temperature 

\begin{equation}\label{Eq1}
T_U ={{\hbar a} \over {2\pi ck_B}}\  ,
\end{equation} 
as measured by a uniformly accelerated observer with proper acceleration $a$. 
This seemingly paradoxical picture 
merely reflects the fact that the particle content of a quantum field theory is 
observer dependent \cite{Ful,Dav}. 

If the acceleration of the object is not uniform, the radiation experienced by it has a frequency 
dependent temperature, whose value is of the same order of magnitude as the one 
given by Eq. (\ref{Eq1}) \cite{Unruhc,BL}.

Since extreme accelerations of about $10^{20}$ meter/sec$^2$ are required in order to produce 
quantum Unruh radiation of temperature $1K$, the effect has no practical implications in 
everyday physics. 
However, we point out that the extremely high accelerations (forces) required for the effect to be significant 
may be found in elementary systems. For example, the quarks inside a nucleon are dominated 
by strong forces, confined within small volumes. 
This raises the possibility that the Unruh effect may have 
significant implications in such elementary systems. 

The primary purpose of this paper is to calculate the energy 
of the Unruh radiation experienced in these systems. We shall assume that in the leading order the 
accelerated particles are semi-classical, with well defined acceleration and position. 
Thus, we only obtain an order of magnitude estimate for the contribution of the Unruh effect in these systems.

The typical temperature (and energy) of the Unruh radiation experienced by an accelerated particle of mass $m$ 
is given by Eq. (\ref{Eq1}), $T_U \sim \hbar a \sim \hbar F/m$. (We use units in which $c=k_B=1$ henceforth.) 
The central binding force $F$ acting on the particle can be approximated by $F \sim {\cal E}' \sim {\cal E}/{\cal L}$, 
where ${\cal E}$ and ${\cal L}$ are the characteristic energy and length-scale of the system, respectively, 
and $'$ denotes a differentiation with respect to the displacement distance from the central equilibrium point. 
The Unruh radiation experienced by the particle is therefore characterized by a temperature  
 
\begin{equation}\label{Eq2}
T_U \sim \hbar {{\cal E} \over {{\cal L}m}}\  .
\end{equation}
The corresponding radiation pressure is given by the Boltzmann formula $P=\pi^2T^4/45\hbar^3$. 

Since the force (hence $T_U$) acting on the {\it finite} size particle varies with distance from the 
central equilibrium point, the particle would detect a radiation {\it gradient}. 
Hence, it is subjected to a buoyant force, caused by the radiation pressure gradient, 
just as an object submerged in a fluid is buoyed up by the non-uniformity of the ambient pressure. 
(For the role of Unruh's radiation buoyant force in black-hole physics, see \cite{Beken,UW} and references therein.) 
The net Unruh force $F_U$ extracted on a particle of characteristic size $d$ 
(due to the pressure gradient $\Delta P$ over an interval $d$) is

\begin{equation}\label{Eq3}
F_U \sim \sigma \Delta P \sim \sigma dP' \sim \sigma dT^3_UT'_U/\hbar^3\  ,
\end{equation}
where $\sigma$ is the characteristic scattering cross section. 
Substituting $T_U$ from Eq. (\ref{Eq2}), and using ${\cal E}' \sim F$, one obtains the ratio

\begin{equation}\label{Eq4}
F_U/F \sim {{\sigma d\hbar{\cal E}^3} \over {{\cal L}^4m^4}}\  .
\end{equation}
The typical size $d$ of an elementary particle is of the order of its Compton length $\hbar/m$, 
implying

\begin{equation}\label{Eq5}
F_U/F \sim {{\sigma \hbar^2 {\cal E}^3} \over {{\cal L}^4m^5}}\  .
\end{equation}
This is the characteristic ratio between the quantum radiation force and the leading-order 
binding force acting on the particle. 
We now provide an order-of-magnitude estimate for this ratio in atomic and nuclear systems. 

{\it Atomic systems.---} The typical length-scale and energy of an atomic system are given by the 
Bohr radius ${\cal L} \sim {\hbar^2}/me^2$, and by its 
Rydberg energy ${\cal E}\sim e^2/{\cal L} \sim {\alpha}^2m$, respectively, 
where $\alpha=e^2/\hbar$ is the fine-structure constant. 
The Unruh temperature (energy) is given by Eq. (\ref{Eq2}), $T_U \sim {\alpha}^3m$. 
Thus, the characteristic energy of the Unruh radiation is much smaller than the binding energy of 
the electron, implying that the scattering cross section is well described by the 
Rayleigh formula (see e.g. \cite{Jac,Sak}), $\sigma \sim R^2_T {\alpha}^4$, where $R_T=e^2/m$ is the Thomson radius. 
Taking cognizance of Eq. (\ref{Eq5}), we find

\begin{equation}\label{Eq6}
F_U/F \sim {\alpha}^{16}\  ,
\end{equation}
Thus, the effect of Unruh radiation in atomic systems is negligible as compared 
to the leading order Coulomb interaction of the particles \cite{Note1}.

{\it Nuclear systems.---} For a quark confined within a nucleon, the characteristic energy and size 
of the system can be approximated by the quark's mass ${\cal E} \sim m$, and by the 
Compton length ${\cal L} \sim \hbar/m$, respectively. Thus, the characteristic temperature (energy) of 
the Unruh radiation in this case is $T_U \sim m$ [see Eq. (\ref{Eq2})]. 
While the exact scattering cross section $\sigma$ is a function of the scattered energy, it may be approximated well 
by the square of the Compton length $\sigma \sim {(\hbar/m)}^2$, see e.g. \cite{Sak}. 
Substituting these in Eq. (\ref{Eq5}), one finds

\begin{equation}\label{Eq7}
F_U/F \sim 1\  .
\end{equation}
This implies that the net Unruh force acting on a quark may have 
an important contribution to the system's dynamics. 

{\it Discussion.---} The quantum Unruh effect is one of the most intriguing phenomenon in modern physics. 
In practice, however, extremely high accelerations (forces) are required in order for the radiation 
to have a significant contribution to the energy and dynamics of a system. 
We point out that such strong forces may exist in nuclear systems, 
in which high energy interactions are confined within small volumes. 
Motivated by this observation, we have calculated the energy of the Unruh radiation in atomic and subatomic systems. 
It was shown that the characteristic energy of the quantum radiation experienced by an elementary particle 
in nuclear systems may be of the same order of magnitude as the particle's own mass. 
The results raise the possibility that Unruh's quantum radiation 
may be of importance to our understanding of the exact structure and dynamics inside a nucleon.

We emphasize that our calculations are semi qualitative-- 
a more quantitative analysis is required in order to determine the exact contribution 
of the Unruh radiation in these systems. 
In particular, it would be important to give both 
the radiation and the particles a full quantum treatment. 
The results presented in this short note are, nevertheless, intriguing, and should motivate further 
study of the Unruh effect in elementary systems.

It is remarkable that the Unruh radiation, whose theoretical discovery was motivated by the 
physics of {\it black holes}, may find important implications in the physics of {\it elementary particles}.

\bigskip
\noindent
{\bf ACKNOWLEDGMENTS}
\bigskip

I would like to thank Jacob D. Bekenstein for helpful discussions. 
This research was supported by G.I.F. Foundation.

\end{document}